\documentstyle[12pt,aasms4,psfig]{article}

\received{ }
\accepted{ }
\journalid{337}{12 May 1998}
\articleid{11}{14}

\slugcomment{Accepted for publication in \apj (Letters)}

\begin{document}

\title{A ``horizon adapted" approach to the study
of relativistic accretion flows onto rotating black holes}

\author{Jos\'e A. Font}
\affil{Max-Planck-Institut f\"ur Gravitationsphysik,
Albert-Einstein-Institut \\
Schlaatzweg 1, 14473 Potsdam, Germany}

\author{Jos\'e M$^{\underline{\mbox{a}}}$. Ib\'a\~nez}
\affil{Departamento de Astronom\'{\i}a y Astrof\'{\i}sica\\
Universidad de Valencia,
Dr. Moliner 50, 46100 Burjassot (Valencia), Spain}

\and

\author{Philippos Papadopoulos}
\affil{Max-Planck-Institut f\"ur Gravitationsphysik,
Albert-Einstein-Institut \\
Schlaatzweg 1, 14473 Potsdam, Germany}

\begin{abstract}
  We present a new geometrical approach to the study of accretion
  flows onto rotating black holes. Instead of Boyer-Lindquist
  coordinates, the standard choice in all existing numerical
  simulations in the literature, we employ the simplest example of a
  {\em horizon adapted coordinate system}, the Kerr-Schild coordinates.
  This choice eliminates unphysical divergent
  behavior at the event horizon. Computations of Bondi-Hoyle accretion
  onto extreme Kerr black holes, performed here for the first time,
  demonstrate the key advantages of this procedure. We argue it offers
  the best approach to the numerical study of the, observationally,
  increasingly more accesible relativistic inner region around black
  holes.
\end{abstract}


\section{Motivation}

Advances in satellite instrumentation, e.g., the Rossi X-Ray Timing
Explorer (RXTE), and the Advanced Satellite for Cosmology and
Astrophysics (ASCA), are greatly stimulating and guiding theoretical
research on accretion physics. The recent discovery of kHz
quasi-periodic oscillations (QPOs), extends the frequency range over
which these oscillations occur into timescales associated with the
innermost regions of the accretion process (for a review see van der
Klis 1996). Stella and Vietri (1997) have proposed that observed low
frequency QPOs in neutron star X-ray binaries correspond to the
precession of the accretion disk, i.e., the Lense-Thirring
effect. This could be the first evidence for a genuinely general relativistic
effect, i.e., the dragging of inertial frames, in such systems.
Morgan, Remillard and Greiner (1997) identified a 67 Hz QPO in the
black hole candidate GRS 1915+105 which may be associated with
relativistic trapped inner disk oscillations (Nowak et al. 1997).
Recently, Cui, Zhang and Chen (1998) have extended the interpretation
of Stella and Vietri to black hole binaries.  Within this model, GRO
J1655-40 and GRS 1915+105 are found to spin at a rate close to the
maximum theoretical limit.
Moreover, in extragalactic sources, spectroscopic evidence (broad iron
emission lines) increasingly points to (rotating) black holes being
the accreting central objects (Tanaka et al. 1995; Kormendy and
Richstone 1995). Recently, Bromley, Miller and Pariev (1998) placed a
limit on the inner edge of the accretor giving rise to the iron
K$\alpha$ emission in MCG-6-30-15 at about 2.6 Schwarzschild
radii. Their estimate is that the black hole is rotating at a rate
which is about $(23 \pm 17)$ \% of the allowed maximum.

Early theoretical studies indicated that a rotating
black hole in the presence of an accretion disk must be spinning at
nearly the maximal rate (Thorne 1974). Rotation increases the
available energy in the near-horizon region: the binding energy per
unit mass of a test particle can reach up to $0.42c^2$ and the
innermost stable circular orbit approaches the horizon (Bardeen, Press
and Teukolsky 1972) and coincides with it (at least in areal
coordinates) in the extreme case $a=M$ ($M$ is the mass of the hole
and $a$ its specific angular momentum). In the rotating case, motions
away from the equatorial plane are affected by the dragging of
inertial frames, while motions in the ergo-region may (more
speculatively) extract rotational energy from the hole. 

Accretion theory is primarily based on the study of stationary flows and
linearized perturbations thereof. Establishing the nature of flow
instabilities, though, will almost certainly require highly resolved
and accurate, time-dependent, non-linear numerical
investigations. Especially intriguing is the possibility of
establishing features of the accretion process that are reflecting the
nature of the spacetime. Such numerical probes rely crucially on
adequate and consistent approximations of the geometry.
For a wide range of accretion problems, a Newtonian theory of gravity
{\em is} adequate for the description of the background gravitational
forces. The extensive experience with Newtonian astrophysics
suggests that the first explorations of the relativistic regime
could benefit from the use of {\em model potentials} (Paczy\'nski and
Wiita 1980). This constitutes the {\em Newtonian paradigm}, which is
still being developed (e.g., Nowak and Wagoner 1991; Artemova,
Bj\"ornsson and Novikov 1996).  For comprehensive numerical work, a
full (i.e., three-dimensional) formalism is required, able to cover
also the maximally rotating hole. In rotating spacetimes the
gravitational forces cannot be captured fully with scalar potential
formalisms. A vivid example is provided by the wave systems examined in
Chandrashekhar (1983), in which rotation introduces {\em frequency}
dependent potentials. Additionally, geometric regions such as the
ergo-sphere would be very hard to model without a metric
description. Whereas the bulk of emission occurs in regions with
almost Newtonian fields, only the observable features attributed to the inner
region may crucially depend on the nature of the spacetime. 

Pioneering numerical efforts in the study of accretion onto black
holes (Wilson 1972; Hawley, Smarr and Wilson 1984; Hawley 1991),
established the relativistic framework, the so-called
{\it frozen star paradigm} of a black hole. In this, the time
``slicing" of the spacetime is synchronized with that of asymptotic
observers far from the hole. This is a powerful approach, leading to a
very economical description of the geometry and can, {\em in
principle}, be used to capture all the interesting effects of the
spacetime curvature. We focus here on the limitations of this approach which
further motivate this {\em Letter}. The shortcomings are due to the
poor choice of coordinates near the black hole horizon and, hence,
manifest themselves only in its neighborhood.  We have argued, though,
that this is precisely the region of most interest. A set of
consistency problems arises from the need for correct boundary
conditions at the horizon. Such conditions are easily imposed on simple supersonic
inflows, but become murkier for co-rotating accretion disks on
rapidly-rotating Kerr black holes. In addition,
imposing boundary conditions on magnetic fields is problematic.
Addressing this issue has led to the development
of the so-called {\it membrane paradigm}. Starting with the description 
of the black hole processes in a non-singular
coordinate system (Damour 1978), this approach endows an {\em
approximate} horizon (a timelike worldtube) with special electric and
magnetic properties and then reverts back to the frozen star picture
for the description of the rest of the spacetime (Thorne and MacDonald
1986). However, the computation is still performed in the original 
singular system. Imposing boundary conditions near the horizon becomes a
demanding practical task, as the singular coordinate coverage of the
horizon leads to an unphysical blowup of coordinate dependent
quantities.

\section{Proposal}

In Papadopoulos and Font (1998) (PF hereafter) we put forward the 
idea of numerically computing
accretion flows onto a black hole in a coordinate system adapted to the
horizon. There, we used simple spherically symmetric flows
to establish its basic feasibility. In addition, we computed
axisymmetric relativistic hydrodynamic accretion onto {\em
Schwarzschild} black holes. We noticed that even the stiff adiabatic
index case $\gamma=2$ was computable, illustrating one of the
advantages of a regular coordinate system. In the remainder of this
{\it Letter} we demonstrate, using a model calculation, the
functionality of our framework to study accretion flows onto
rapidly-rotating black holes.

Coordinate systems attached to physical observers are generically
regular at the horizon, but lack the important practical property of
stationarity. If, besides regularity, we impose 
the additional requirements of a stationary metric and
spacelike foliation we obtain what we call {\em horizon adapted
coordinates}. A comprehensive family of those
systems for the Kerr spacetime can be obtained with the following
transformation from the standard Boyer-Lindquist (BL)
$(t,r,\theta,\phi)$ to the new coordinates
$d\tilde{\phi} = d\phi + (a/\Delta) dr$ and 
$d\tilde{t} = dt + \left[ \frac{1+Y}{1+Y-Z} - \frac{1-Z^k}{1-Z}\right] dr$,
where $Y= a^2 \sin^2\theta/\varrho^2$, $Z=2Mr/\varrho^2$,
$\varrho^2 =r^2+a^2\cos\theta^2$, $\Delta=r^2-2Mr+a^2$,
and $k$ is a non-negative integer which parametrizes the family
(natural units are used throughout,
$G=c=1$).  All members of the family are regular at the horizon, hence
the algebraically simplest choice ($k=1$) is preferred.  This
corresponds to the so-called Kerr-Schild (KS) form of the Kerr metric.
With this choice the line element becomes
\begin{eqnarray}
ds^2 &=& - (1-Z) d\tilde{t}^2
          - 2 a Z\sin^2\theta d\tilde{t}d\tilde{\phi} 
          + 2 Z d\tilde{t} dr \nonumber  \\
          && + (1+Z) dr^2   
          - 2a (1+Z)\sin^2\theta dr d\tilde{\phi} 
          + \varrho^2 d\theta^2  \nonumber  \\
          && + \sin^2\theta [{\varrho^2} 
                + a^2(1+Z) \sin^2\theta] d\tilde{\phi}^2.
\label{ksform}
\end{eqnarray}
The regularity at the horizon, which is located at the largest root of
the equation $(\Delta=0)$, is manifest. A more extensive discussion,
in particular of the wider range of choices available in the
non-rotating case, is given in PF.

The central aspect of our computational paradigm is that the integration
domain includes, in a natural way, the event horizon of the black
hole. In fact, it extends {\em inside} the horizon and can be
truncated at an arbitrary inner radius. This prescription solves, at
once, both conceptual and practical problems in the near horizon
region. Boundary conditions for fields can be imposed in an
unambiguous manner, whereas there is a sharp reduction in the spatial
gradients of the solution. The first benefit arises from the nature of
the horizon as a one way (ingoing) membrane. In the absence of good
boundary conditions even ``poor" conditions would be acceptable,
provided they allow a stable and converging computation. For example,
unphysical reflections or numerical heating will never re-emerge from the
horizon, since the characteristic speeds for such processes are
necessarily slower than $c$. The reduced resolution requirements arise
from the fact that the projections of four-vectors onto the
three-dimensional integration space is everywhere regular, given the
regular coordinate system. 

The regularization of the horizon introduces
two new non-zero metric elements. However, this additional algebraic
complexity should not be much of a concern for relativistic
integration algorithms which must be designed to handle a general
metric. On the other hand, it is a fact that considerable intuition
and mathematical tools have been obtained in the simpler frozen star
form of the Kerr metric. This background work can still be used by
transforming geometric quantities back and forth.
For stationary accretion patterns this
process is entirely straightforward and we include a demonstration
of such transformations below.

\section{Accretion onto a rapidly-rotating black hole}

The generic astrophysical scenario in which matter is accreted in a
non-spherical way by a compact object is the one suggested originally
by Bondi and Hoyle (1944). The astrophysical importance of this
process has led to its detailed numerical investigation
(see e.g., Benensohn, Lamb and Taam 1997 and references therein).
Recently, the relativistic version of the Bondi-Hoyle accretion has
been extended to the non-axisymmetric, non-rotating black hole case
(Font and Ib\'a\~nez 1998b). Its generalization to account for
rotating black holes remains still unexplored. We take here the first
step in that direction, investigating the problem adopting a simple
geometrical scenario. At the same time, this scenario offers an
adequate test-bed calculation to demonstrate the proposal of horizon
adapted coordinates for accretion onto {\em rotating black holes}.

In performing the computation, we adopt the ``infinitesimally thin"
accretion disk setup. This is motivated by simplicity considerations,
before attempting three-dimensional studies. This initial setup has
been used to some extent in Newtonian simulations of wind accretion in
cylindrical and Cartesian coordinates (see, e.g., Matsuda et al 1991,
Benensohn, Lamb and Taam 1997 and references therein).  In our setup
we use a restricted set of equations, where the vertical structure of
the flow is assumed not to depend on the polar coordinate. This
assumption requires that in the {\em immediate neighborhood} of the
equator, vertical (polar) pressure gradients, velocities and gravity
(tidal) terms vanish. Those conditions are strictly correct at the
equator for flows that are reflection symmetric there. Hence, despite
the obvious reduction of generality, important insights into the
accretion process can be obtained in a simplified framework. In
particular, as will be apparent from the discussion of the flow
morphology below, our dimensional simplification still captures the
most demanding aspect of the Kerr background, which is encoded in the
large azimuthal shift vector near the horizon. This is most relevant
for the issues discussed in this work.

For the computation we use the general relativistic hydrodynamic
equations as presented in Banyuls et al. (1997). Hence we take advantage
of the explicit knowledge of the characteristic fields in order to
build up a linearized Riemann solver to handle discontinuous
solutions. The formulation of
the equations in Schwarzschild coordinates and the numerical algorithm
can be found in Font and Ib\'a\~nez (1998a). Tests of the code can be
found in Font et al (1994) (for the subclass of special relativistic
flows) and Banyuls et al (1997) (for general relativistic flows in
Schwarzschild backgrounds). To perform the simulation we present
below, we specialize the equations to the KS form of the metric,
Eq.~(\ref{ksform}). Specific details, including the description of the
time-dependent variables, fluxes and source terms of the equations in
BL and KS coordinates, as well as a comparative study of the accretion
flows in those systems will be reported in a forthcoming paper (Font,
Ib\'a\~nez and Papadopoulos 1998b; FIP hereafter).

The initial state of the fluid is completely characterized by the
asymptotic conditions upstream the accretor. We choose as free
parameters the asymptotic velocity $v_{\infty}=0.5$, the sound speed
$c_{s_{\infty}}=0.1$ and the adiabatic index $\gamma=5/3$. In
addition, as the accretor is a rotating black hole, we have an extra
parameter, $a$, its angular momentum per unit mass. We choose a
rapidly-rotating hole with $a=0.999M$. The
simulation is performed in the equatorial plane
($\theta=\pi/2$) using an ($r$, $\tilde{\phi}$) grid of $200 \times 160$ zones.

Representative results of our simulations are plotted in Fig.~1, where
we show isocontours of the logarithm of the density at the final time
$t=500M$. The hole is rotating counter-clockwise.  This simulation
employs KS coordinates, with the innermost grid radius placed at $1.0
M$, i.e., {\it inside} the horizon, $r_+\equiv M+\sqrt{M^2-a^2}$,
which for our model is $1.04M$. The outermost radius corresponds to
$50M$. The dotted line indicates the position of the horizon.
The simulation is characterized by
the presence of a well-defined tail shock. The flow pattern reaches a
steady-state around $t\approx 100-200M$, which was confirmed computing
the time evolution of the mass accretion rate as described in FIP.

The transformations to horizon adapted coordinates are non-trivial,
from a geometric point of view, in the sense that a new {\em time}
coordinate is involved. There is no one-to-one correspondence between
snapshots of evolutions in the two different 
systems. Still, for accretion flows that relax to a stationary state,
the time dependence is, by definition, factored out and a direct
comparison is possible by transforming four dimensional tensor
quantities appropriately (details are given in FIP). 
Making use of the final stationarity of the flow, we plot in Fig.~2
how the accretion pattern would look like were the computation
performed in BL coordinates. The transformation induces a noticeable
wrapping of the shock around the central hole. The shock would wrap
infinitely many times before reaching the horizon, due to
an additional coordinate pathology of the BL system.
As a result, the computation in BL coordinates, although in 
principle possible, would be much more challenging than in KS 
coordinates, particularly near to the horizon. Genuine effects, such as frame
dragging are expected in the case of accretion not on the equatorial
plane. Since the last stable orbit approaches closely the horizon in
the case of maximal rotation, the interesting scenario of co-rotating
extreme Kerr accretion would be severely affected by the strong
gradients which develop in the strong-field region.
This will most certainly affect the accuracy and,
potentially, also the stability of numerical codes.  An example of
instabilities possibly related to coordinate issues has been reported
recently (Igumenshchev and Beloborodov 1997).

\section{Summary}

We have shown the feasibility of a new geometrical approach to the
numerical study of accretion flows onto rotating black holes.
Our procedure relies on the use of {\it horizon adapted coordinate
systems} in which all fields, i.e., metric, fluid and electromagnetic
fields, are free of coordinate singularities at the event
horizon. Among the large family of those systems
we propose the use of the Kerr-Schild coordinate system -- the
simplest example of the class -- as the natural framework to perform
accurate numerical studies of relativistic accretion flows.  We have
discussed our approach in the context of the various paradigms in
black hole astrophysics. Our proposal shares with the frozen star
picture the exact representation of the relativistic geometry. It
departs from it in significant ways in the crucial horizon region, in
which a choice of different time coordinate allows the explicit use of
the ``one way membrane'' picture in the computations.  We conclude
that smooth coordinate systems at the horizon become an invaluable
tool for the numerical study of accretion phenomena around extreme
Kerr black holes. Coupled with high resolution numerical methods these
systems may help clarify the basic dynamical processes around
accreting black holes.

\section*{Acknowledgements}

It is a pleasure to thank Hara Papathanassiou for suggestions that
significantly improved the manuscript.  This work is partially
supported by the Spanish DGICYT (grant PB94-0973).  J.A.F acknowledges
financial support from a TMR European contract (nr. ERBFMBICT971902).
All computations were performed at the AEI.

\newpage

\begin{figure}
\centerline{\psfig{figure=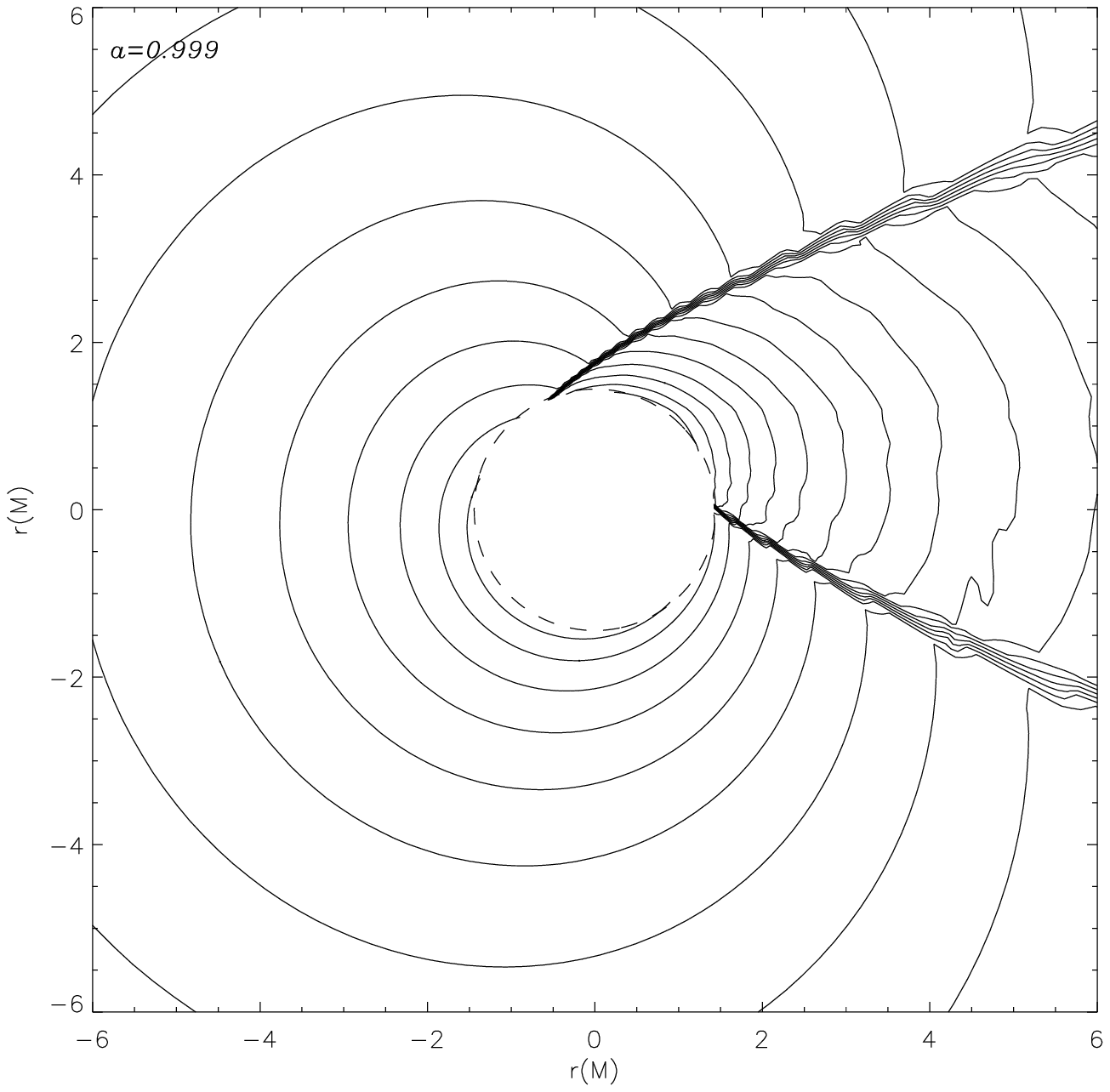,width=5.0in,height=5.0in}}
\caption[Figure 1.]{Relativistic 
Bondi-Hoyle accretion onto a rapidly-rotating black hole ($a=0.999M$)
in KS coordinates. 
We plot 20 isocontours of the logarithm of the density
(scaled to its asymptotic value), ranging from
$-0.13$ to $2.37$, at the final stationary time $t=500M$. 
Asymptotically, the flow moves from left to right.
The dashed line indicates the location of the horizon.}
\end{figure}

\newpage

\begin{figure}
\centerline{\psfig{figure=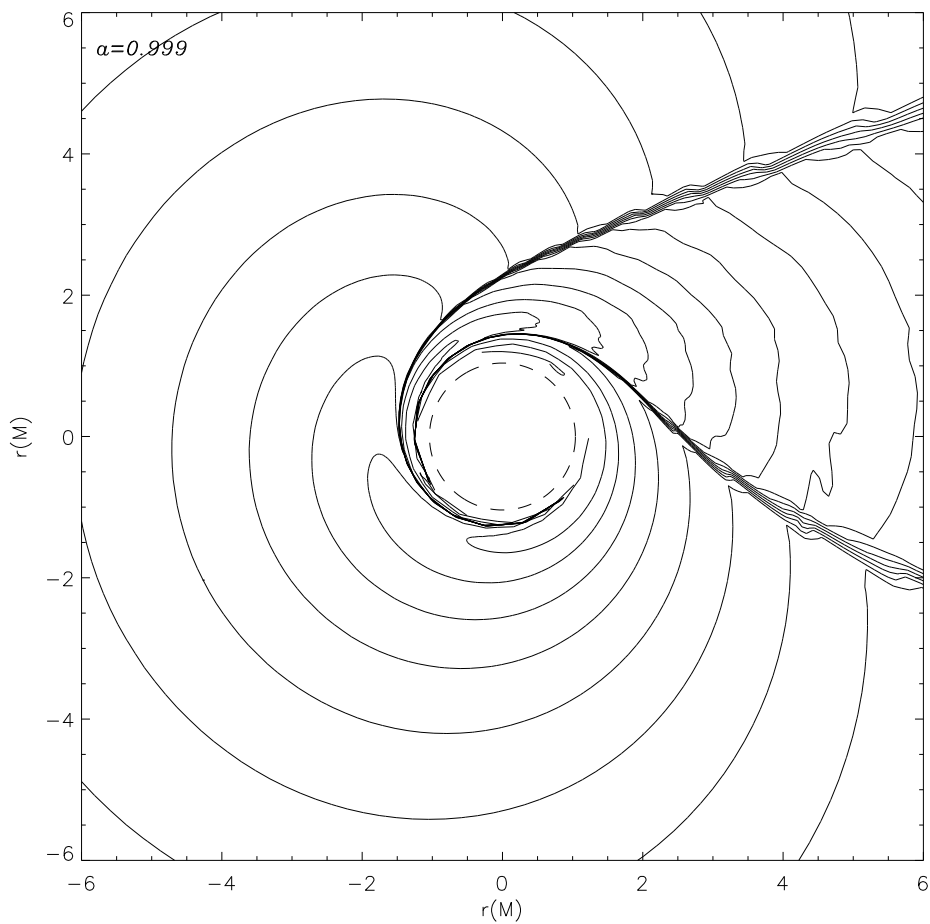,width=5.0in,height=5.0in}}
\caption[Figure 2.]{
Transformation of the stationary accretion
pattern of Fig.~1 to BL coordinates. There are 20 isocontours
of the logarithm of the density spanning the interval $-0.17$ to $2.28$.
The shock appears here totally
wrapped around the horizon. It does not extend
all the way down to the horizon as the coordinate
transformation is singular there.}
\end{figure}

\end{document}